\documentclass[review]{elsarticle}

\usepackage{lineno,hyperref}
\usepackage{amssymb}
\usepackage{lineno}
\usepackage{booktabs}
\usepackage{multirow}
\usepackage[]{algorithm2e}
\usepackage{comment}
\usepackage{subfig}
\usepackage[numbers]{natbib}
\usepackage{caption}
\usepackage{hyperref}
\usepackage{etoolbox}
\usepackage[utf8]{inputenc}

\modulolinenumbers[5]

\journal{Journal of Computer Methods and Programs in Biomedicine}

\patchcmd{\pprintMaketitle}
 {\hrule}
 {\clearpage\hrule}
 {}{}
\appto\endfrontmatter{\clearpage}

\begin{document}

\begin{frontmatter}

\title{Integration of Convolutional Neural Networks for Pulmonary Nodule Malignancy Assessment in a Lung Cancer Classification Pipeline}

\author[1]{Ilaria Bonavita\fnref{contrib}}
\ead{ilaria.bonavita@eurecat.org}

\author[1,2]{Xavier Rafael-Palou\fnref{contrib}\corref{cor1}}
\ead{xavier.rafael@eurecat.org}

\author[2]{Mario Ceresa\corref{cor2}}
\ead{mario.ceresa@upf.edu}

\author[2]{Gemma Piella\corref{cor2}}
\ead{gemma.piella@upf.edu}

\author[1]{Vicent Ribas\corref{cor2}}
\ead{vicent.ribas@eurecat.org}

\author[2,3]{Miguel A. González Ballester\corref{cor2}}
\ead{ma.gonzalez@upf.edu}

\fntext[contrib]{Both authors contributed equally in this work.}
\cortext[cor1]{Corresponding author}

\address[1]{Eurecat, Centre Tecnològic de Catalunya, eHealth Unit, Barcelona, Spain}
\address[2]{BCN Medtech, Dept. of Information and Communication Technologies, Universitat Pompeu Fabra, Barcelona, Spain}
\address[3]{ICREA, Barcelona, Spain}

\begin{abstract}
\textbf{Background and Objective:} The early identification of malignant pulmonary nodules is critical for a better lung cancer prognosis and a less invasive chemo or radio therapies. Nodule malignancy assessment done by radiologists is extremely useful for planning a preventive intervention but is, unfortunately, a complex, time-consuming and error-prone task. This explains the lack of large datasets containing radiologists malignancy characterization of nodules; \textbf{Methods:} In this article, we propose to assess nodule malignancy through 3D convolutional neural networks and to integrate it in an automated end-to-end existing pipeline of lung cancer detection. For training and testing purposes we used independent subsets of the LIDC dataset; \textbf{Results:} Adding the probabilities of nodules malignity in a baseline lung cancer pipeline improved its F1-weighted score by 14.7 \%, whereas integrating the malignancy model itself using transfer learning outperformed the baseline prediction by 11.8 \% of F1-weighted score; \textbf{Conclusions:} Despite the limited size of the lung cancer datasets, integrating predictive models of nodule malignancy improves prediction of lung cancer. 


\end{abstract}


\end{frontmatter}


\section{Introduction}
\label{intro}
Lung cancer is the uncontrolled growth of abnormal cells in one or both lungs. These abnormal cells can form tumors and interfere with the normal functioning of the lung, which provides oxygen to the body via the blood.

Although the 5-year survival for lung cancer has improved over the last fifty years, it is still one of the most common cancers, accounting for over 225,000 cases, 150,000 deaths, and \$12 billion in health care costs yearly in the U.S. \cite{bib:choi2013automated}. It is also one of the deadliest cancers; only 17\% of people in the U.S. diagnosed with lung cancer survive five years after the diagnosis, and the survival rate is even lower in developing countries. 

Early detection of lung cancer significantly improves the chances of patient survival. However, in most cases, a patient is unaware that she/he has a pulmonary nodule until a chest X-ray or a low-dose computed tomography (CT) scan of the lungs is performed. For this reason, early stage detection of benign and malignant pulmonary nodules plays an important role in clinical diagnosis. Today, the gold standard for lung cancer detection consists in routinely taking a CT scan, and detecting nodules (i.e. small and approximately spherical masses) in it. Once lung nodules are detected, radiologists perform size measurements to assess malignancy. To support them in this task, several guidelines like LungRADs \cite{american2014lung} and Fleischner  \cite{bib:macmahon2005guidelines} have been proposed. These guidelines are a compilation of well documented cases and a set of rule-based recommendations from the clinical experience designed to help clinicians to discern among pulmonary nodules, normal tissues and artifacts, as well as to determine the inherent malignancy of the nodules. However, they are constrained to a limited number of visual parameters (e.g. size, morphology, texture and location of the nodules) and to a fixed range of values.

 Low-dose CT is an effective method for radiologists to early identify lung cancer \cite{bib:national2011reduced}, although it presents several limitations. First, radiologists need to process large volumes of CT slices, usually with a low signal-to-noise ratio, which causes erroneous classifications of regions with weak or irregular contours. In addition, lung cancer diagnosis through CT is often subjective and highly affected by observer's experience, fatigue and emotional state \cite{bib:patz2014overdiagnosis}, which can lead to inconsistent results from the same radiologist at different times or from different radiologists examining the same CT image.

Emulating the decision process of radiologists to determine malignancy of a nodule would be an extremely useful tool to help physicians plan future interventions for patients. Several approaches can be found in the literature relying on artificial intelligence and computer vision techniques. Conventional solutions (e.g. \cite{kaya2015weighted, gonccalves2017learning}) propose engineering handcrafted features extracted directly from the CT image to build standard machine learning classifiers. This approach achieves satisfactory results when nodule candidates are well-defined, but shows some shortcomings when the nodules present complex and different sizes, shapes and context. An alternative recent solution to this problem is the use of deep convolutional neural networks (e.g. \cite{causey2018highly,shen2015multi}) that are able to learn automatically inherent representations directly from the raw images. 

In this work, we use 3D deep convolutional neural networks to build accurate malignancy classifiers using annotations made by radiologists on pulmonary nodules. The main contribution of this paper with respect to previous works is two-fold. First, we provide a framework to allow integrating nodule malignancy classifiers, built at nodule level, into a pipeline that does not take into account malignancy information, but predicts lung cancer at the patient level. To this aim, three different types of integration were designed: using the predicted classes, the probabilities or the models themselves. Secondly, we quantified the contribution of the nodule malignancy classifiers for lung cancer prediction. For this objective we evaluated the three different types of integration and we compared their performances with that of a baseline lung cancer pipeline.

The paper is organized as follows: in the next section we review the existing related work on nodule malignancy and cancer classification. Then we present the methods and materials used. Finally we provide the results and a thorough discussion on the main outcomes presented in the article.

\section{Related Works}

In the past years numerous works have addressed the problem of classifying the malignancy of pulmonary nodules in CT scans; some of these works use as features only radiologists annotations of the nodules and perform classification for example with rule-based \cite{kaya2015weighted} and statistical learning \cite{Hancock2016LungNM} methods or by building a machine learning classifier \cite{gonccalves2017learning} or classifiers ensemble \cite{Zinovev2009PredictingRP, Zinovev2011BuildingAE}. 
\\
In other works, in addition or as alternative to radiologist annotations, shape-based, margin-based, and texture-based features \cite{dhara2016combination} or 3D features of the nodules \cite{reeves2016automated} are computed directly from the image with classical image analysis techniques.  
\\
In more recent years, it has been shown that deep learning techniques can outperform standard techniques in discriminating benign from malignant nodules (e.g. \cite{shen2015multi, xie2018fusing, song2017using}). In \cite{hua2015computer} a deep belief network is used to extract from nodules features that are fed to a convolutional neural network aimed at classifying the nodule malignancy. In \cite{kumar2015lung} deep features are extracted from an autoencoder. In \cite{causey2018highly} high malignancy classification accuracy is achieved by using a convolutional neural network and radiological quantitative features.

Despite the abundance of papers focusing on classification of nodule malignancy and on nodules detection in CT scans, little effort has been put in providing a systematic analysis of the effects of combining both to answer the question of whether predicting malignancy at a nodule level is beneficial for cancer prediction at patient level. To the best of our knowledge, \cite{shen2016learning} is the closest work to ours that tackles this question. However, the focus of \cite{shen2016learning} is limited to the transferability of deep features of nodules to the cancer prediction task and the input data are exactly located nodules. Our aim is, instead, to provide and evaluate different types of nodule malignancy integration within an end-to-end cancer detection pipeline that takes as input raw CT scans.

\section{Materials and Methods}

\label{S:2}
\subsection{The Data}
\subsubsection{LIDC and LUNA16 datasets}\label{S:2.1}
The LIDC \cite{armato2015data} is the largest publicly available reference database for lung nodules. It contains a total of 1018 CT scans each of which associated with a file containing annotations from four experienced thoracic radiologists. The annotations are the result of a two-phase reading process in which the radiologists were asked to mark suspicious lesions and to provide additional characterization of lesions of diameter larger or equal to 3 mm marked as nodule \cite{armato2011lung}. 
\\
In this work we use an updated version of the LIDC dataset provided in the LUNA16 challenge \cite{setio2017validation}, which includes only scans with at least one lesion of size $>=$ 3 mm marked as nodule by at least three of the four radiologists. The LUNA16 dataset consists of 888 CT scans comprising a total of 1186 nodules. Annotations with coordinates of each nodule in the three spatial axes inferred from the original LIDC annotations are also provided. 
\\
We obtained the malignancy outcome of our classifiers from the annotation files in the LIDC database as they provide, among other characteristics, the subjective assessment of each radiologist of the likelihood of malignancy of the nodule. The admitted malignancy scores are discrete values ranging from 1 (highly unlikely for cancer) to 5 (highly suspicious for cancer).
Since for each nodule included in LUNA16 we have the assessment of three or four radiologists, in order to obtain a unique label we averaged their scores.

\subsubsection{TCIA Diagnosis Data}
For 130 cases the LIDC dataset provides diagnostic data at patient level obtained from biopsy, surgical resection, progression or reviewing of the radiological images showing nodules stable after two years \cite{clark2013cancer}. 
\\
We retained this small dataset from the data used for building the malignancy classifiers, and we used it for training and testing the baseline and integrated lung cancer classifiers.

\subsection{Method}

\subsubsection{The Malignancy Classifiers}\label{sec:mal_classifier}
\begin{figure}%
    \centering
    \subfloat[]{{\includegraphics[height=5.6cm]{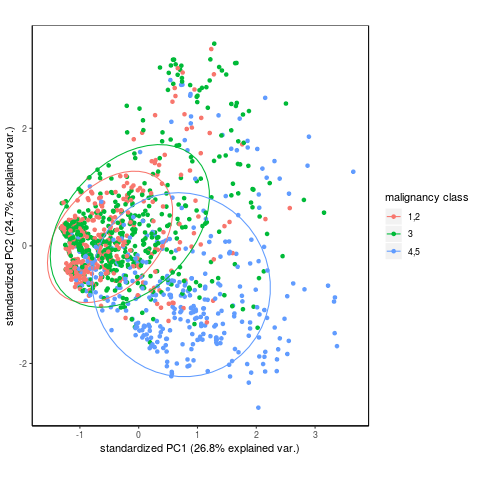}}}%
    \qquad
    \subfloat[]{\raisebox{0.3cm}{\includegraphics[height=5cm]{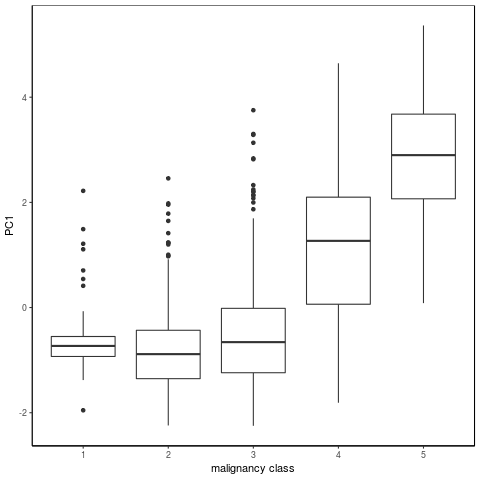}}}%
    \caption{PCA analysis and boxplot of radiologists annotated features per malignancy class.}%
    \label{fig:pca_mal}%
\end{figure}

We describe here the approach used to build the nodule malignancy classifiers that will be integrated in the cancer prediction pipeline.
\\
The input data to train the malignancy classifier consists of 3D cubes measuring (32, 32, 32) mm centered in the centroid of the nodule computed from the coordinates in the LUNA16 annotation file. Note that each CT scan (i.e. subject) can contain more than one nodule; hence, to avoid any data leakage we assigned all the nodules belonging to a subject to only one of the training, validation or test sets. Additionally, we performed clipping (using a filter of [-1000, 400] HU) and normalization of the cubes.
\\
The malignancy score of each nodule was obtained from the original XML annotations using a parser provided by the second place winner of the DSB Kaggle competition \cite{hammack2017forecasting} and averaging the radiologist scores as described in \ref{S:2.1}. 
\\
Given the binary nature of the final cancer prediction we want to provide, we decided to remove nodules of ambiguous or intermediate malignancy from our experimental dataset. A Principal Component Analysis performed on some of the most relevant features annotated by radiologists showed that nodules of malignancy 1, 2 and 3 have similar feature distributions differently from those of malignancy 4 and 5 (Figure \ref{fig:pca_mal} b). Additionally, nodules of class 3 present higher variance, and form a less well defined cluster in the principal components space (Figure \ref{fig:pca_mal} a). We decided therefore to remove them from our analysis.
We hence opted for training and validating our classifiers on: \textsl{Dataset\_145}, in which we selected only nodules labelled as 1, 4 or 5 and \textsl{Dataset\_1\&245}, in which we selected nodules of malignancy 4 and 5 and we merged in one single category (renamed 1\&2) nodules labelled as 1 and 2. 
Both datasets were split in training (60\%) and validation (40\%) sets in a stratified fashion. As stated above, the test set for both the malignancy classifiers and the cancer pipeline consists of the TCIA data. However, only CTs containing at least one nodule with label 1, 4 or 5 (\textsl{Test\_145}) or 1,2,4 or 5  (\textsl{Test\_1\&245}) were selected. 
Sample sets size and labels distribution are presented in Table \ref{tbl:lab_distr}.

\begin{table}[htb!]
\centering
\begin{tabular}{llllll}
\hline
\multirow{2}{*}{Dataset} & \multirow{2}{*}{N subjects} & \multicolumn{4}{c}{N nodules} \\ \cline{3-6} 
                         &                             & 1(\&2)  & 4    & 5   & total  \\ \hline
Dataset\_145             & 247                         & 72      & 213  & 48  & 333    \\
Dataset\_1\&245          & 351                         & 287     & 213  & 48  & 548    \\
Test\_145                & 65                          & 15      & 65   & 9   & 89     \\
Test\_1\&245             & 82                          & 59      & 65   & 9   & 133   
\end{tabular}
\caption{Dataset used for building and testing the malignancy classifier.}
\label{tbl:lab_distr}
\end{table}

The 3D nodule-cubes and corresponding malignancy labels are fed to a machine learning multi-class classifier. The model is based on a deep learning convolutional neural network (CNN). We tested two different networks: a 3D CNN with 3 convolutional layers, each followed by a 3D Max-Pooling layer and with a final dropout layer (\textsl{shallowCNN}), and a similar 3D CNN (\textsl{deeperCNN}) to which a dropout and a batch normalization layers were added for each convolutional layer. 


\begin{figure}[htb!]
\centering
  \includegraphics[width=0.95\textwidth]{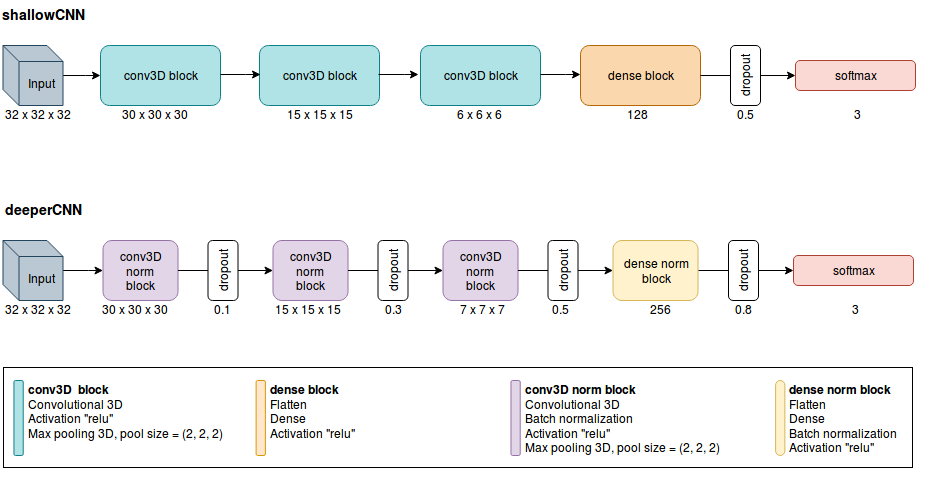}
  \caption{Architecture of the two CNN networks used for the malignancy classifiers.}
  \label{fig:cnn_blocks}
\end{figure} 

We trained the model weights with a batch training approach for 150 epochs and adopting early stopping with Adam optimizer \cite{kingma2014adam} for regularization. We set the learning to 0.001 and we chose categorical cross-entropy as loss function. Moreover, given the small size of our dataset, we used data augmentation on the training set (90 degrees of rotation, 0.02 of shear, zoom range of 0.1, shift of 0.05 and horizontal and vertical flip).
\\
Different training and validation batch sizes together with other input parameter combinations have been tested. A detailed description of the network architectures is presented in Figure \ref{fig:cnn_blocks}.
 \\
The combination of the two network architectures and of the two datasets led to the creation of four malignancy classifiers:  \textsl{shallowCNN\_145}, \textsl{shallowCNN\_1\&245}, \textsl{deeperCNN\_145},  \textsl{deeperCNN\_1\&245}.
\\
Performances of the four classifiers are presented in results section whereas the integration of the classifiers and the evaluation in the cancer prediction pipeline is described below.

\subsubsection{The lung cancer pipeline}
\label{cancer_pipeline}

With the intention of setting a baseline method, we developed a two-stage lung cancer pipeline that did not take into account any information regarding nodule malignancy. We refer to this pipeline as our baseline method.
\\

\begin{enumerate}
\item \textit{Nodule detection}

To build the automatic nodule detection stage, we used the LUNA16 dataset (reserving 10\% for testing purposes) since it contains, for each CT, location and diameter of the nodules. The first process performed was re-sampling each CT to an isotropic resolution (1, 1, 1) mm in order to reduce the variance given by the different pixel size/coarseness (e.g. the distance between slices) of the scans.

Secondly, we performed a segmentation of the lungs from the re-sampled CTs, with the intention of reducing the analysis to the area of interest. For this task, we relied on a method proposed by the most cited kernel of the Data Science Bowl Kaggle competition \cite{bib:lungSeg}. This method consists in applying a threshold (i.e. -320 HU) to separate the air from the tissues. Then, it uses connected components to separate the lung air from outside, and finally it applies a morphological dilation to fill the existing gaps in the lung tissue.

To detect nodule candidates in a CT, we used a 3D blob detector based on the Difference of Gaussian method \cite{bib:valente2016automatic}. This technique tries to detect nodules by retrieving those parts of the image that differ in properties, such as brightness or grey-level, compared to surrounding regions. One advantage of this method is its intuitive parameterization. In particular, we needed to tune 5 parameters: the minimum and maximum diameter of the region to look for (i.e. the minimum and maximum Gaussian standard deviations), the steps (i.e. the number of standard deviations to try between the defined ranges), a similarity threshold and the overlap score used for pruning closely located regions of interest. The configuration selected for this method was 5 mm and 60 mm as minimum and maximum nodule diameters, 5 steps, a threshold of 0.15 and 0.9 of overlapping. More details on the evaluation results of this method are available in the supplementary material.

As this candidate detection method tends to be optimistic (i.e. to accept several candidates similar in shape and texture to nodules), we implemented a classifier aimed at reducing the rate of false positive candidates. We chose to solve this task with a 3D CNN and, after empirical tests with different network architectures, we opted for the ResNet-50 \cite{bib:he2016deep}. To train this network, we used the same training set used for building the nodule detection method, along with a list of candidate node locations, provided by the LUNA16 challenge. Inputs of the network were volumes of (32, 32, 32) mm extracted from the nodule candidate positions. We used 0.0001 as initial learning rate, Adam optimization and binary cross-entropy for the loss function. Additionally, to improve the generalization ability of the network, in the training phase we used data augmentation of the positive class by a factor of 1:240. In particular, we applied 90 degree of rotation, 0.2 of shear, zoom range of 0.1, up and downs shifts of 0.5 and horizontal and vertical flips. The network reached its best performance in training phase after 6 epochs with a batch size of 32. Further details regarding the evaluation of this method are also available in the supplementary material.

\item \textit{Cancer classification}\label{sec:canc_classifier}

The following stage of the pipeline consisted in building a lung cancer classifier, fed with the detected nodules, in order to predict cancer probability for each patient. For this purpose, we used the TCIA dataset that provides only patient labels (cancer or non-cancer). Given the lack of nodule labels, one of the main difficulties we had to face in building the classifier was to establish a nodules-patient labels relationship. We created a ground truth for the detected nodules from the ground truth of the patients by labelling all the nodules detected in a CT as 0/1 depending on the presence (1) or absence (0) of cancer in the patient scan. For example, if three nodules were detected by the pipeline in a CT scan of a patient with cancer, all the nodules were labelled as cancerous. Thus, we constructed a lung cancer classifier that predicts the probability of cancer of every nodule in a CT. Then, since we were interested to report cancer predictions at the patient level, we reported as cancer probability of the patient the predicted cancer probability of his/her most cancerous nodule (i.e. the highest among the predicted cancer probabilities of all his/her nodules). 

Additionally, in the classification we included the main features provided by the 3D blob detector. In total we selected three main features (radius, power and relative\_z\_position) referring respectively to size, intensity and location of the nodules. Although further image descriptors could be envisaged, we limited our choice to those three not only to highlight the contribution of the nodule malignancy knowledge but also to approximate as closely as possible the features recommended in the current radiologist guidelines to focus on when screening nodules in a CT scan.

Several classification algorithms were used to train the classifiers, each accounting for a different classification strategy (i.e. linear, non-linear, distance-based, and tree-based). Moreover, different hyper-parameters were defined for each algorithm (Table S5 of the supplementary material). In order to determine the best classification model, we used a grid-search 5-fold cross-validation, a technique suitable for our sample size range \cite{friedman2001elements}. 

\end{enumerate}

\subsubsection{The Nodule Malignancy Integration}

\begin{figure}[ht!]
\centering
  \includegraphics[width=1.0\textwidth]{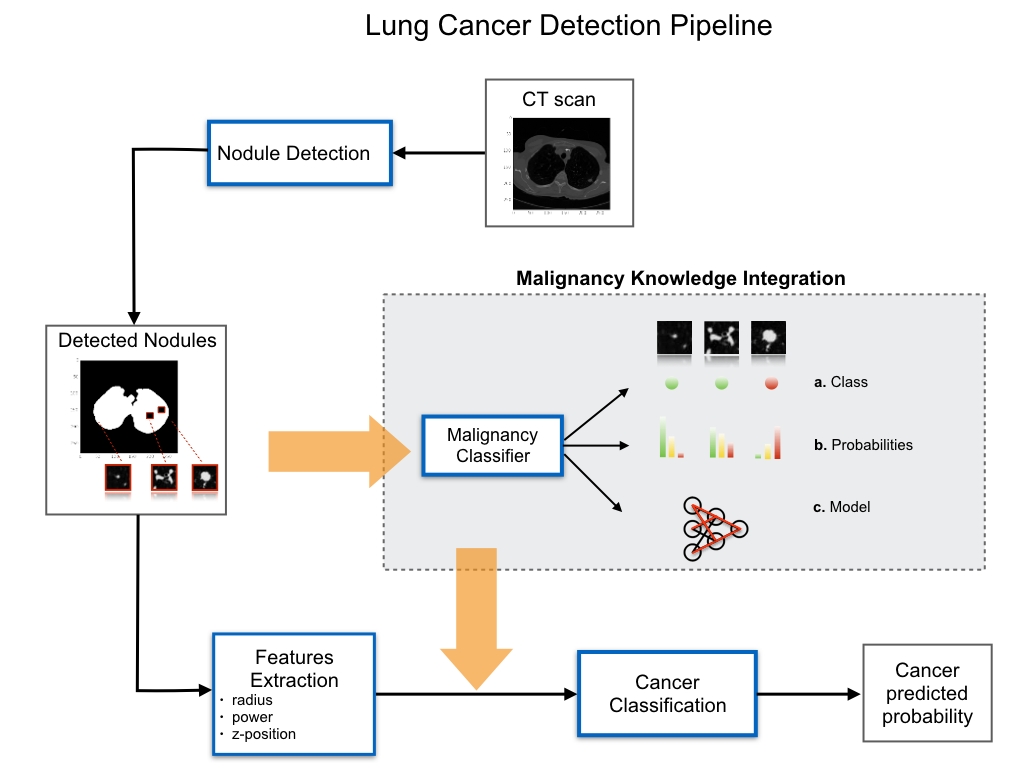}
  \caption{Pipeline proposed for lung cancer classification.}
  \label{fig:pipeline}
\end{figure}

In order to assess the effects of the automatic nodule malignancy classification (section \ref{sec:mal_classifier}) for lung cancer prediction, we proposed three different methods to integrate the nodule malignancy models in the lung cancer pipeline: integration of predicted classes, integration of probabilities or integration of the models themselves (Figure \ref{fig:pipeline}).

\begin{enumerate}
\item Class integration

The integration method using classes consisted in creating a new categorical feature containing the label predicted by the nodule malignancy classifiers. Thus, this feature was 0, 1 or 2 depending on whether the malignancy classifier predicted malignancy level of 1 ( or 1\&2 for \textsl{shallowCNN\_1\&245} and \textsl{deeperCNN\_1\&245} classifiers), 4, 5 respectively. To build the lung cancer classifier, we then concatenated this feature to the three basic features defined in  \ref{cancer_pipeline} (cancer classification), namely, radius, power and z-position.
\\

\item Probability integration

The second integration method consisted in creating three new features, each containing the predicted probability of the nodule to be of malignancy level 1 ( or 1\&2for \textsl{shallowCNN\_1\&245} and \textsl{deeperCNN\_1\&245} classifiers), level 4  or level 5. To build the lung cancer classifier, we then concatenated these three features to the three basic features. 

\item  Model integration

The third integration method aimed to directly use the nodule malignancy models for the task of lung cancer prediction. Several techniques can be envisaged for this type of integration. We proposed using transfer learning \cite{goodfellow2016deep} since both problems have the same type of input data (CT scans) and a similar objective (identifying malignancy). To perform transfer learning, all the weights of the layers of the 3D malignancy networks were frozen, the last softmax layer was removed and replaced by a dense network (several configuration parameters of this network are presented in the supplementary material) and a final sigmoid layer. The first layer of the dense network was combined with the three basic features defined for the lung cancer classifier of the pipeline. The last layer of the final network outputs a value between 0 and 1 that represents the probability of lung cancer. 
\end{enumerate}

For tuning and evaluating the classifiers, independently of the integration method used, we applied grid-search and 5-fold cross-validation as we did for building the cancer classifier of the pipeline. 

\section{Results}
\label{S:3}

\subsection{Malignancy Classification Results}
We present here the results of the nodule malignancy classifiers. Although nodule classification is not the focus of our work, it is important to determine that these classifiers are able to extract useful information from the CTs before integrating them in the cancer pipeline. In Table \ref{tbl:mal_results} we summarize the weighted average performance metrics and the macro averaged F1-score on the test set of the four classifiers. The models \textsl{shallowCNN\_145} and \textsl{deeperCNN\_145} achieved best performances with batch size of 32 in training and validation, while for \textsl{shallowCNN\_1\&245} and \textsl{shallowCNN\_1\&245} batch size of 32 and 16 respectively in training and validation were selected. 
In all the experiments we augmented each nodule in the training set by a factor between 10 and 25, augmenting more nodules of malignancy 5 given their lower representation in the dataset.

Overall, the more shallow architectures slightly outperformed the deeper ones; nevertheless, all the classifiers achieved weighted F1-score above 0.75 with the best one (\textsl{shallowCNN\_1\&245}) achieving 0.83. These results indicate that the nodule deep features extracted by the CNN are good predictors of nodule malignancy.

\begin{table}[]

\centering
\caption{Results of nodule malignancy classification on test set (at nodule level).}
\begin{tabular}{@{}llllll@{}}
\toprule
Classifier     & Precision & Recall & F1-score & F1-macro & Support \\ \midrule
shallowCNN\_145 & 0.83      & 0.81   & 0.82    & 0.68 & 89     \\
deeperCNN\_145   & 0.80       & 0.73   & 0.76     & 0.63 & 89      \\
shallowCNN\_1\&245 & 0.83      & 0.83   & 0.83     & 0.67 & 133     \\
deeperCNN\_1\&245    & 0.82      & 0.80    & 0.81     & 0.66 & 133     \\ \bottomrule
\end{tabular}
\label{tbl:mal_results}
\end{table}

\subsubsection{Consistency between nodule-level malignancy predictions and patient-level diagnostic ground truth}
To validate our hypothesis that the integration of a nodule malignancy classifier in a cancer detection pipeline can improve the predictions, we evaluated the consistency between the diagnosed cancer status of a patient and the predicted malignancy of his/her nodules. To do so, we inferred the cancer label of each patient from the malignancy labels of his/her nodules: if the CT scan of the patient contains at least one nodule with predicted malignancy 4 or 5, then the patient is positive to cancer, otherwise (i.e. all the nodules in the CT are benign) the patient is negative to cancer. Given this rule, we obtained cancer predictions at patient level in the cases where the predictions of nodule malignancy come from: 1) the radiologists, 2) the four malignancy classifiers. Performance metrics of these rule-based predictions are evaluated in the \textsl{Test\_145} and \textsl{Test\_1\&245} sets (as they are the only provided with truth cancer labels) and are reported in Table \ref{tbl:ct_mal_compare}. It is worth noticing that both radiologist and CNN classifiers achieved comparably high, although not perfect, predictions (in Test\_145 the best F1-score was 0.92 achieved by radiologists and  \textsl{deeperCNN\_145} while in Test\_1\&245 the best F1-score was 0.85 achieved by \textsl{shallowCNN\_1\&45} followed by 0.84 obtained from the radiologists prediction).

\begin{table}[h!] 
\caption{Cancer prediction at patient level from nodule malignancy.}
\centering
\begin{tabular}{@{}llllll@{}}
\toprule
Dataset         & Prediction source & Precision & Recall    & F1-score  & Support \\ \midrule
Test\_145    & radiologist       & 0.89 & 0.94 & 0.92 & 65      \\
Test\_145    & shallowCNN\_145       & 0.86 & 0.96 & 0.91 & 65      \\
Test\_145    & deeperCNN\_145         & 0.88 & 0.96 & 0.92 & 65      \\ \midrule
Test\_1\&245 & radiologist       & 0.89 & 0.79 & 0.84 & 82      \\
Test\_1\&245 & shallowCNN\_1\&245       & 0.87 & 0.84 & 0.85 & 82      \\
Test\_1\&245 & deeperCNN\_1\&245         & 0.83 & 0.78 & 0.80 & 82      \\ \bottomrule
\end{tabular}
\label{tbl:ct_mal_compare}
\end{table}

\subsection{Lung cancer results}

The pipeline described in section \ref{cancer_pipeline} was applied on the diagnosed TCIA dataset. From the 130 cases, we obtained that 100 (76.9\%) were predicted with potential lung nodules, 11 cases (8.4\%) were correctly predicted without any cancerous nodule and 19 cases (14.1\%) were false negatives as they had some missing cancerous nodules.

\begin{figure}[ht!]
\centering
  \includegraphics[width=1.0\textwidth]{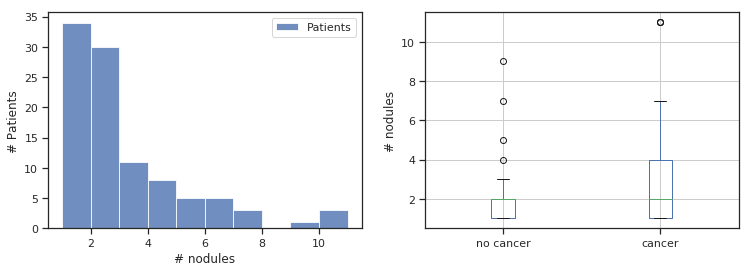}
  \caption{Data distribution for lung cancer classification.}
  \label{fig:nodules}
\end{figure}

On the 100 CT cases with detected nodules (227 nodules), we ran the cancer classification stage of the pipeline. The data was imbalanced with a non-cancer/cancer class ratio of 1:3.61. This ratio was respected during the random partitioning of the data in training and test datasets. In total, for training we had 75 cases (21 non-cancer, 54 cancer) with 220 nodules (48 non-cancer, 172 cancer). In contrast, for testing we had 25 cases (6 non-cancer, 19 cancer) with 57 nodules (12 non-cancer, 45 cancer). Figure \ref{fig:nodules} shows the distribution of nodules by patient and the box-plot of nodules for cancer and non-cancer CTs.

The results of evaluating the different malignancy integration pipelines for lung cancer prediction are summarized in Table \ref{tbl:intRes}. This table shows the weighted precision, recall and F1-scores for cross-validation, test at the nodule level and test at the patient patient level. The baseline method achieved 0.65 +/- 0.021 of weighted F1 in cross-validation, whereas 0.55 in test at the nodule level and 0.593 in test at the patient level. The pipeline with malignancy probabilities integration method achieved the best results with 0.709 of weighted F1 in test at the nodule level and 0.74 of F1-weighted score in test at the patient level.

\begin{figure}[ht!]
\centering
  \includegraphics[width=1.0
  \textwidth]{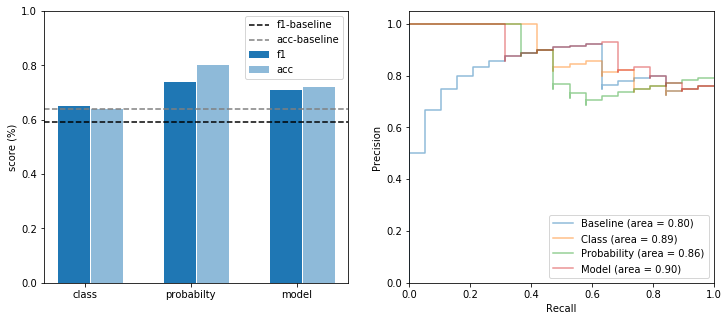}
  \caption{Performance comparison of the lung cancer pipelines.}
\label{fig:mal_filter_compare}
\end{figure}

Figure \ref{fig:mal_filter_compare} shows a bar-plot with the accuracy and the weighted F1-scores achieved by the different integration pipelines. The dashed lines represent the baseline classification performances. On the right, we show a precision-recall curve of the different lung cancer pipelines. This curve is especially appropriate when the classes are imbalanced as it shows the trade-off between precision and recall for different thresholds \cite{bib:saito2015precision}. 

\begin{table}[ht!]
\centering
\caption{Cross-validation and test (ND: nodule level, PT: patient level) results for the lung cancer pipelines.}
\begin{tabular}{llllll}
\hline 
\multirow{2}{*}{} & \multirow{2}{*}{Metric} & \multirow{2}{*}{\begin{tabular}[c]{@{}l@{}}Baseline\\ Pipeline\end{tabular}} & \multicolumn{3}{c}{\begin{tabular}[c]{@{}c@{}}Malignancy Integrated\\ Pipelines\end{tabular}} \\ \cline{4-6} 
 &  & & Class & Probabilty & Model  \\ \hline 
\multirow{3}{*}{CV}
 & prec  & 0.627+/-0.03                                             & 0.737+/-0.01              & 0.766+/-0.02  & 0.715+/-0.06 \\ 
 & rec                      & 0.711+/-0.05                       & 0.587+/-0.03               & 0.732+/-0.03   & 0.712+/-0.05  \\ 
  & F1                      & 0.650+/-0.02                                               &0.623+/-0.02             & 0.743+/-0.02 & 0.712+/-0.05              \\ \hline
\multirow{3}{*}{Test (ND)}   
 & prec                    & 0.615                      & 0.685                 & 0.692    & 0.703   \\ 
 & rec                   & 0.509                                                & 0.491                  & 0.737     & 0.684  \\ 
 & F1                    & 0.55                                                & 0.536                 & 0.709    & 0.693   \\ \hline  
\multirow{3}{*}{Test (PT)}   \\
 & prec                    & 0.553                      & 0.66                 & 0.842    & 0.704   \\ 
 & rec                   & 0.64                                                & 0.64                  & 0.8     & 0.72  \\ 
 & F1                    & 0.593                                                & 0.64                 & 0.74    & 0.711   \\ \hline  
\end{tabular}
\label{tbl:intRes}
\end{table}

\section{Discussion}
\label{S:4}

One of the most critical tasks that radiologists have to perform when examining lung CTs is to identify nodules from normal lung tissue. Highly malignant nodules are usually candidates of being lung cancer, therefore radiologists should precisely quantify the malignancy of the pulmonary nodules before planning expensive and sometimes traumatic clinical interventions. 

Measuring nodule malignancy is a complex and tiresome process with significant levels of intra- and inter-observer variability. Several tools relaying on image processing and conventional machine learning techniques or, more recently, deep convolutional neural networks have been proposed to support radiologists in this task. However, to the best of our knowledge, very few of them ( e.g. \cite{shen2016learning}), independently of the technique selected, use nodule malignancy for the classification of lung cancer. With the intention of providing a realistic evaluation of the importance of nodule malignancy for the automatic lung cancer classification, in this work we have provided a framework with different methods to integrate nodule malignancy in a cancer detection pipeline. 

To this aim, we created several nodule malignancy classifiers using 3D convolutional neural networks. To build these classifiers, beforehand, we knew the level of malignancy, the position and the size of the nodules to classify. The best nodule malignancy classifier (shallowCNN\_1\&245) achieved a relevant performance 0.83 of weighted F1-score when classifying the malignancy of the nodules in an independent test set. 

The expected usefulness of these classifiers to the task of lung cancer prediction was validated by deriving a cancer classification from the nodule malignancy prediction on the TCIA diagnosed dataset. The best malignancy classifier (deeperCNN\_145) achieved a performance of 0.92 of weighted F1 score, comparable to the performance using the malignancy annotations given by the radiologists. However, it is worth noting that the evaluation was performed knowing a priori the location of the nodules and that the nodules annotated with a label 3 were removed due to their ambiguous malignancy. 

To have a more realistic evaluation, we first created a baseline pipeline comprising a nodule detection and a cancer classification that uses a very simple set of descriptors (such as the radius, intensity and location of the candidates). We limited the number of features to this basic set to reasonably emulate the features recommended in the current radiologist guidelines. Also, since our primary objective was not to offer high cancer classification performance but to quantify the importance of nodule malignancy for lung cancer, reducing the number of features allowed us to decrease the training time and thus increase the number of experiments. 

Eventually, to assess the effects of automatic nodule malignancy classification for lung cancer prediction, we provided three different ways to integrate the nodule malignancy classifiers into a lung cancer pipeline. The first approach aimed to use only the predicted classes as a new feature to add into the basic set of features of the baseline pipeline. The second approach consisted in creating three new features, representing the nodule malignancy probability distribution, and adding them to the features of the baseline. Finally, the last integration method consisted in using directly the malignancy model for lung cancer classification. In particular, we used a transfer learning technique which consisted on freezing the weights of the malignancy classifiers, removing the last layer and replacing it by new dense layers.

In total three new pipelines were created by applying the different integration techniques within the baseline pipeline. The three pipelines and the baseline were trained using the TCIA dataset and evaluated using grid-search with a 5-fold cross-validation. 

Results show that the best pipeline with integrated nodule malignancy outperforms up to a 15.9\% and 14.7\% of weighted F1 score in comparison with the baseline at the nodule and patient level. The best pipeline was using the malignancy probabilities and it achieved a difference of 2.9\% of weighted F1 score at the patient level with respect to the second best integration pipeline, the malignancy model integration.
This result may appear surprising since the model integration adds to the classifier more features and hence more information. However, this extra information comes at the cost of an increased dimensionality of the problem, suggesting that this transfer learning approach may be better suited when a larger dataset would be available. Alternatively, a further fine tuning (e.g. unfreezing or removing more layers) of the transfer learning proposed can be envisaged.
Nevertheless, the model integration pipeline significantly outperformed the baseline by 11.8\% of weighted F1 score at the patient level. In contrast, malignancy class integration did not significantly improve the lung cancer classification performance of the baseline. The poorer performance of the class compared to the other integration methods was expected, since the information was compressed in a single categorical feature not able to capture the complexity of the problem.

The findings of our study suggest that systematically integrating the assessment of nodule malignancy in an automated cancer detection system may improve significantly the ability of the system to identify cancer in lung scans. Emulating the malignancy assessment with powerful techniques such as deep learning, able to extract complex information directly from raw data, can relieve the difficulties and costs of a manual assessment. However, we believe that the lack of larger datasets with manual malignancy annotations and diagnostic cancer labels constitues the main limitation of our study. If in the future datasets of this kind become available, our pipeline will highly benefit from the additional amount of information, which will likely result in more accurate predictions. Better predictions will eventually: reduce the need for time-consuming manual annotations and feature engineering approaches, provide a reliable support to radiologists and automatize to a greater extent cancer detection pipelines adopted in clinical applications.

Our work is, to the best of our knowledge, the first attempt to build this nodule-malignancy/patient-cancer integrated framework. Despite the encouraging results, several improvements can be envisaged to extend this approach. For instance, creating an ensemble of all the malignancy classifiers rather than using them individually could enhance the classification performance. Furthermore, nodule malignancy could be also used for filtering nodule candidates detected by the cancer pipeline. Thus, rather than using all the detected nodules, we could use only the most malignant ones as input for the lung cancer classifier.

\section{Conclusions}
In this study we have proved that it is feasible to build highly accurate malignancy classifiers relying on deep learning techniques to predict nodule malignancy. We have validated that they are also good predictors of lung cancer at the patient level when having the location of nodules beforehand. In order to provide a more realistic evaluation of nodule malignancy for lung cancer classification, we finally proposed a novel framework to quantify and assess nodule malignancy for lung cancer given only CTs and labels at the patient level.  
The experimental findings of our study suggest that systematically integrating the assessment of nodule malignancy in an automated cancer detection system improves up to 14.7\% of F1-score the ability of the system to identify cancer in lung scans. The encouraging results presented are, to the best of our knowledge, the first attempt to build this nodule-malignancy/patient-cancer integrated framework to quantify nodule malignancy for future research in lung cancer classification.

\section*{Conflict of interest}

The authors confirm that there are no conflicts of interest associated with this publication and there has not been significant finantial support for this work that could have influenced its outcome.

\section*{Acknowledgement}

The authors want to acknowledge Filip Velickovski for its valuable work in the implementation of the nodule detection methods of the lung cancer pipeline. 
This work was partially funded by the Industrial Doctorates Program
(AGAUR), grant number DI079.

\section*{References}

\bibliography{main}

\end{document}


\renewcommand{\thepage}{S\arabic{page}} 
\renewcommand{\thesection}{S\arabic{section}}  
\renewcommand{\thetable}{S\arabic{table}}  
\renewcommand{\thefigure}{S\arabic{figure}} 

\title{Supplementary material}
\author{}
\date{}

\maketitle



\section{The lung cancer pipeline}

Here we present the results of the first stage of the lung cancer pipeline. Those were obtained using an independent testset (10\% of the data) of the LUNA16 dataset.

\subsection{Nodules detection}

Tables \ref{tbl:config_nd} and \ref{tbl:res_nd} show the description and results of three different configurations tested for the nodule detection part of the lung cancer pipeline.

\begin{table}[htb!]
\centering
\begin{tabular}{l|l|l|l|}
\cline{2-4}
                                     & Option 1 & Option 2 & Option 3 \\ \hline
\multicolumn{1}{|l|}{Minimum radius} & 10       & 5        & 5        \\ \hline
\multicolumn{1}{|l|}{Maximum radius} & 40       & 30       & 60       \\ \hline
\multicolumn{1}{|l|}{Steps}          & 10       & 10       & 5        \\ \hline
\multicolumn{1}{|l|}{Threshold}      & 0.2      & 0.15     & 0.15     \\ \hline
\multicolumn{1}{|l|}{Overlap}        & 0.9      & 0.7      & 0.9      \\ \hline
\end{tabular}
\caption{Configurations of the Difference of Gaussian method for lung nodules detection.}
\label{tbl:config_nd}%
\end{table}

\newpage

\begin{table}[htb]
\centering
\begin{tabular}{l|l|l|l|}
\cline{2-4}
                                                                                                                       & \multicolumn{3}{c|}{DoG configurations}                                                                                                                                                                                                                         \\ \cline{2-4} 
                                                                                                                       & \begin{tabular}[c]{@{}l@{}}Option 1\end{tabular} & \begin{tabular}[c]{@{}l@{}}Option 2\end{tabular} & \begin{tabular}[c]{@{}l@{}}Option 3\end{tabular} \\ \hline
\multicolumn{1}{|l|}{Total detected nodules}                                                                           & 21                                                                                 & 29                                                                                   & 73                                                                                  \\ \hline
\multicolumn{1}{|l|}{Total detected candidates}                                                                        & 1142                                                                               & 7130                                                                                 & 76631                                                                               \\ \hline
\multicolumn{1}{|l|}{\begin{tabular}[c]{@{}l@{}}Min,Max,Mean,Std radius \\ of detected nodules (real)\end{tabular}}    & \begin{tabular}[c]{@{}l@{}}3.51 \\ 12.14 \\ 8.03 \\ 2.33\end{tabular}              & \begin{tabular}[c]{@{}l@{}}2.8 \\ 12.14 \\ 6.82 \\ 2.74\end{tabular}                 & \begin{tabular}[c]{@{}l@{}}1.7 \\ 12.14 \\ 4.69 \\ 2.65\end{tabular}                \\ \hline
\multicolumn{1}{|l|}{\begin{tabular}[c]{@{}l@{}}Min,Max,Mean,Std radius \\ of detected nodules (predicted)\end{tabular}}    & \begin{tabular}[c]{@{}l@{}}5.0 \\ 12.6 \\ 8.6 \\ 2.31\end{tabular}                 & \begin{tabular}[c]{@{}l@{}}3.05 \\ 12.29 \\ 7.33 \\ 2.74\end{tabular}                & \begin{tabular}[c]{@{}l@{}}2.5 \\ 8.66 \\ 4.41 \\ 2.35\end{tabular}                 \\ \hline
\multicolumn{1}{|l|}{\begin{tabular}[c]{@{}l@{}}Min,Max,Mean,Std intensity \\ of detected nodules (pred)\end{tabular}} & \begin{tabular}[c]{@{}l@{}}0.21 \\ 0.45 \\ 0.31 \\ 0.07\end{tabular}               & \begin{tabular}[c]{@{}l@{}}0.16 \\ 0.57 \\ 0.32 \\ 0.13\end{tabular}                 & \begin{tabular}[c]{@{}l@{}}0.15 \\ 1.31 \\ 0.46 \\ 0.3\end{tabular}                 \\ \hline
\multicolumn{1}{|l|}{Total missing nodules}                                                                            & 84                                                                                 & 76                                                                                   & 32                                                                                  \\ \hline
\multicolumn{1}{|l|}{\begin{tabular}[c]{@{}l@{}}Min,Max,Mean,Std radius \\ of missing nodules (real)\end{tabular}}     & \begin{tabular}[c]{@{}l@{}}1.64 \\ 8.36 \\ 3.2 \\ 1.16\end{tabular}                & \begin{tabular}[c]{@{}l@{}}1.64 \\ 8.36 \\ 3.15 \\ 1.25\end{tabular}                 & \begin{tabular}[c]{@{}l@{}}1.64 \\ 6.28 \\ 2.97 \\ 1.12\end{tabular}                \\ \hline
\end{tabular}
\caption{Results from three different configurations of the Difference of Gaussian method for lung nodules detection. The total number of nodules in the test set was 105.}
\label{tbl:res_nd}
\end{table}

\subsection{False Positive Reduction}

\begin{figure}[htb!]
\centering
  \includegraphics[width=0.95\textwidth]{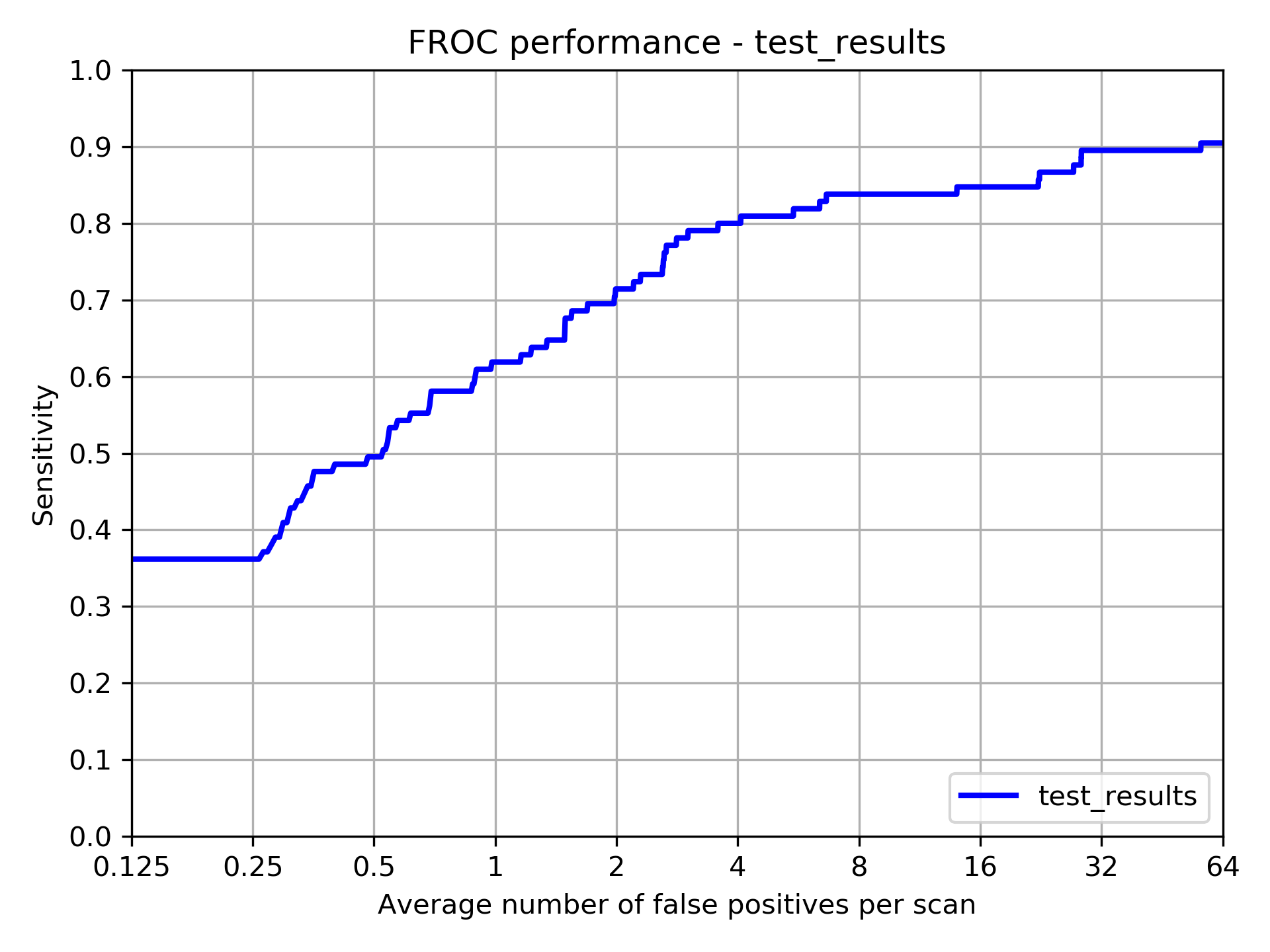}
  \caption{FROC curve achieved in testing for the 3D ResNet network.}
  \label{fig:froc}
\end{figure}

Tables \ref{tbl:res_cm}, \ref{tbl:res_fp} and Figure \ref{fig:froc} present the results achieved by the 3D ResNet deep convolutional network used for the false positive reduction task.

\begin{table}[htb!]
\centering
\begin{tabular}{llll}
                      &                                    & \multicolumn{2}{l}{Predicted}                                                                                                                      \\ \cline{3-4} 
                      & \multicolumn{1}{l|}{Real}          & \multicolumn{1}{c|}{\begin{tabular}[c]{@{}c@{}}False\\ (0)\end{tabular}} & \multicolumn{1}{c|}{\begin{tabular}[c]{@{}c@{}}True\\ (1)\end{tabular}} \\ \cline{2-4} 
\multicolumn{1}{l|}{} & \multicolumn{1}{l|}{Candidate (0)} & \multicolumn{1}{l|}{75726}                                               & \multicolumn{1}{l|}{54}                                                 \\ \cline{2-4} 
\multicolumn{1}{l|}{} & \multicolumn{1}{l|}{Nodule (1)}    & \multicolumn{1}{l|}{58}                                                  & \multicolumn{1}{l|}{86}                                                 \\ \cline{2-4} 
\end{tabular}
\caption{Confusion matrix results for the 3D ResNet network.}
\label{tbl:res_cm}
\end{table}

\begin{table}[htb!]
\centering
\begin{tabular}{l|l|l|l|l|}
\cline{2-5}
                                    & Precision & Recall & F1-score   & Support \\ \hline
\multicolumn{1}{|l|}{Candidate (0)} & 1.00      & 1.00   & 1.00 & 75780   \\ \hline
\multicolumn{1}{|l|}{Nodule (1)}    & 0.61      & 0.60   & 0.61 & 144     \\ \hline
\end{tabular}
\caption{Classification results for the 3D ResNet network.}
\label{tbl:res_fp}
\end{table}

\newpage

\subsection{Cancer classification}

In this section we describe the pipeline parameters (Table \ref{tbl:ml_setup}) used for training the machine learning classifiers as well as the parameters used for the dense fully connected network (Table \ref{tbl:dl_setup}) for lung cancer prediction.

\begin{table}[htb!]
\caption{Pipeline parameters tested using grid-search and 5-fold CV.}
\begin{center}
\begin{tabular}{r@{\quad}ll}
\hline
\multicolumn{1}{c}{\rule{0pt}{10pt}Algorithm}&\multicolumn{2}{c}{Options}\\[2pt]
\hline\hline\rule{0pt}{12pt}
k-NN         
& \begin{tabular}{l}
	n\_neighbors = [1,3,5,7,9,11] \\
    weights = ['uniform', 'distance']
\end{tabular} & \\ \hline
LR 
& \begin{tabular}{l}
C = [0.001,0.01,0.1,0.5,1,3] \\
class\_weight = ['balanced'] \\
penalty = ['l1', 'l2']
\end{tabular} & \\ \hline
RF          
& \begin{tabular}{l}
  n\_estimators = [100,150,200,250,500,750] \\
  criterion = ['entropy','gini'] \\
  max\_depth = ['None',2,4,6] \\
  class\_weight = ['balanced']
\end{tabular} & \\ \hline
SVM  
& \begin{tabular}{l}
  C = [0.001,0.01,0.1,0.5,1,3] \\
  gamma = [0.005,0.01, 0.05,0.1,1,3] \\
  kernel = ['radial','poly'] \\
  degree = [3,5,7,9] \\
  class\_weight = ['balanced'] \\
\end{tabular} & \\ \\[2pt]
\hline
\end{tabular}
\end{center}
\label{tbl:ml_setup}
\end{table}

\begin{table}[htb!]
\caption{Parameters for training the dense network.}
\begin{center}
\begin{tabular}{l|l}
\hline
Method & Options \\[2pt]
\hline\hline\rule{0pt}{12pt}
Hidden-Layers & 
\begin{tabular}{l}
(size/4),(size/3),\\
(size/2),(size) \end{tabular} \\ \hline
Alpha & 1e-5,1e-3,1e-2,1,3,10 \\ \hline
Activation & 'relu', 'sigmoid' \\ \hline
Solver & 'lbfgs' \\ \hline
Max\_iter & 200 \\ \hline
Tol & 1e-4 \\ \hline
\hline
\end{tabular}
\end{center}
(*) The value of 'size' is the output of the N-1 layer of the nodule malignancy model together with the 3 features of the lung cancer baseline pipeline.
\label{tbl:dl_setup}
\end{table}